\documentclass[prb,showpacs,preprintnumbers,twocolumn]{revtex4}
\usepackage{graphicx}
\usepackage{amssymb}
\begin{document}

\title{Soft Mode in cubic PbTiO$_3$  by Hyper-Raman Scattering}

\author{J. Hlinka$^1$, B. Hehlen$^2$, A. Kania$^3$,
 and I. Gregora$^{1}$}

\affiliation{$^1$Institute of Physics, Academy of Sciences of the
Czech Republic, Na Slovance 2, 18221 Praha 8, Czech Republic}
\affiliation{$^2$Laboratoire des Colloides, Verres et Nanomat\'eriaux (LCVN), UMR CNRS 5587,\\
University of Montpellier II, 34095 Montpellier, France}
\affiliation{$^3$Institute of Physics, University of Silesia,
PL-40-007 Katowice, Poland}

\begin{abstract}
Hyper-Raman scattering experiments allowed collecting the spectra
of the lowest $F_{1u}$-symmetry mode of PbTiO$_3$ crystal in the
paraelectric phase up to $\approx 930$\,K as well as down to about
1\,K above the phase transition.
 It is realized that this mode is
fully responsible for the Curie-Weiss behavior of its dielectric
permittivity above $T_{\rm c}$. Near the phase transition, this
phonon frequency softens down to 17\,cm$^{-1}$ and its spectrum
can be well modeled as a response of a single damped harmonic
oscillator. It is concluded that PbTiO$_3$ constitutes a clean
example of a soft mode-driven ferroelectric system.
\end{abstract}

\date{\today}



\maketitle

Since the seminal works of Frohlich, Ginzburg, Cochran and
Landauer,\cite{Froh49,Ginz49a,Ginz49b,Coch59,Land87} it is well
understood that high dielectric constant of many practically
exploited polar materials is directly related to the presence of a
low frequency polar excitation -- through the Lyddane-Sachs-Teller
relationship. In displacive-type ferroelectrics, this excitation
has a character of a vibrational lattice mode, so-called soft
mode. Soft mode frequencies are typically found in 0.1-1 THz
frequency range, so that the high permittivity
 persists throughout the whole MHz-GHz range and it is usually
quite a robust intrinsic property. Other materials may reveal high
static permittivities also for other reasons, for example due to
mobile ferroelectric domain walls, space-limited conductivity at
various structural inhomogeneities etc.\cite{Lunk02,LiuJ04,LiMi09}
As a rule, such mechanisms usually lead to high permittivities
only at very low frequencies and the dielectric properties of
these materials are also much more susceptible to various fatigue
phenomena. Therefore, the THz-range frequency dispersion of the
dielectric permittivity provides the key information needed to
understand the potential of practical use of the high permittivity
materials.

A nice example of a material owing its high permittivity to a
well-defined THz-range frequency polar vibration is the quantum
paralectric SrTiO$_3$.\cite{Bark62,Rupp62} In most of its
ferroelectric counterparts, however, soft polar fluctuations
cannot not be described by a single DHO oscillator.\cite{Petz87}
In fact, it is tempting to conclude that there is
 some kind of additional central mode
in the vicinity of the all ferroelectric phase transitions.
 At least, additional THz range excitations related to the soft
phonon degree of freedom have been quite neatly experimentally
identified in rather different perovskite polar materials, such as
tetragonal ferroelecric BaTiO$_3$\cite{Hlin08} on one hand and in
a pseudocubic relaxor ferroelectric
PbMg$_{1/3}$Nb$_{2/3}$O$_3$\cite{ZeinSM} on the other hand.
Seemingly similar phenomena may probably be of a rather different
nature, but in both cases the presence of the additional
excitation in the dielectric spectrum is crucial for explaining
the magnitudes of the measured static permittivity.

Therefore, it seems also important to provide clean examples of
ferroelectric materials where the THz range permittivity is
actually well described by a single polar soft mode excitation.
Originally, the paraelectric PbTiO$_3$ has been considered as the
example of such a simple soft mode system.\cite{Shir70} The
presence of central peak
 was later proposed to exist there as well, but only in a GHz frequency range.\cite{Font90,Font91,Girs97}
In reality,
  the  measurements of the soft mode spectra above the fairly high ferroelectric phase transition temperature (T$_{\rm C}$=760\,K)
 of PbTiO$_3$ turned out to be technically quite challenging.
 Already the pioneering inelastic neutron scattering (INS) study of Shirane et al.\cite{Shir70} reported that accurate measurements in the vicinity of the Brillouin zone center are difficult and the data reported there are in fact extrapolated from phonon dispersion relations. Similar difficulties have been encountered in subsequent INS experiments,\cite{Kemp06,Tome12} and even less straightforward is the analysis of the confocal Raman light scattering spectra reported in Ref.\,\onlinecite{Soon08}.

 In fact, there are quite sizable disagreements
 among the published estimates for soft mode frequency and its temperature dependence in the paraelectric phase.
  For example, the extrapolation of the soft mode frequency to the phase transition temperature gives about 11\,cm$^{-1}$ according to Ref.\,\onlinecite{Tome12}, 20\,cm$^{-1}$ according to Ref.\,\onlinecite{Shir70}, and more than 60\,cm$^{-1}$ according to Ref.\,\onlinecite{Soon08}.

  The aim of this paper is to report new set of reference spectral data in the vicinity of a ferroelectric phase transition, demonstrating that the paraelectric
PbTiO$_3$ can be indeed considered as such a  clean
case where the THz response can be well modeled by a single DHO.

A clear-cut answer to these issues was obtained here  by means of
 Hyper-Raman scattering (HRS).
Previously, soft modes has been studied by this technique in a
range of polar perovskite materials.
\cite{Pre83,Heh07,Ali08} In principle, state-of-art
instrumental resolution allows measurements of polar modes in the
paraelectric phase by HRS down to a few cm$^{-1}$. Present HRS
experiments have been performed using the setup described in Ref.\,\onlinecite{Ali08}, only upgraded with an optical microscope.

 The sample was a 1\,mm thick platelet of a PbTiO$_3$ single
crystal grown at University of Silesia,\cite{Kani06} with a natural surface
 normal to the [001] ($z$-axis). It was mounted in a hot stage with the
[100] ($x$) and [010] ($y$) crystallographic axes parallel to the
horizontal (H) and vertical (V) direction, respectively. The data
were collected in the backscattering geometry with $q\parallel$
[001]. A broadband half-wave plate followed by a Glan-Thomson
polarizer allowed for analysis of the scattered light polarized
along V or H. A 1800\,grooves/mm grating and a 150\,$\mu$m
entrance slit give rise to a spectral resolution of $\sim$
3\,cm$^{-1}$. Cooling and heating rates did not exceed 10\,K/min
with a
 temperature stabilized within $\sim\pm$1\,K.
Since the strong SHG signal in the ferroelectric phase persists up
to the phase transition $T_{\rm C}$ the soft mode spectra could be
measured only in the paraelectric phase. On the other hand, the
onset of SHG and Raman scattering at $T_{\rm C}$ allowed us to
determine quite precisely the measured temperatures with respect
to $T_{\rm C}$.


The simple $O_h^1$ (Z=1) perovskite structure of paraelectric
PbTiO$_3$ has four zone center optic modes, three $F_{1u}$ polar
modes and one $F_{2u}$ non-polar vibration, all being triply
degenerate and active in HRS. The lowest frequency $F_{1u}$ is
known to play the role of the soft mode. A representative set of
resulting HRS spectra revealing this mode is shown in Fig.\,1. All
these spectra show a broad soft mode response and a
resolution-limited central line. Up to about 50\,K above $T_{\rm
C}$, the broad soft mode gives a bell-like response centered
 at zero energy. Above this temperature the response function clearly shows two broad maxima, corresponding to the Stokes and anti-Stokes spectral bands.

\begin{figure}[ht]
\includegraphics[width=8cm]{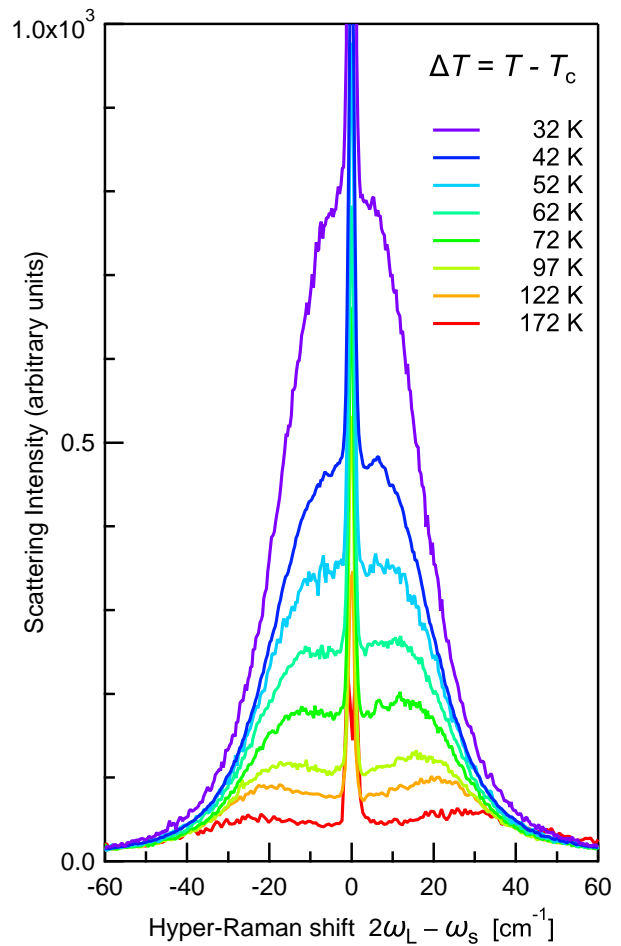}
\caption{(color online) Unpolarized (V+H) HRS spectra of in
paraelectric phase of PbTiO$_3$ single crystal for several
temperatures indicated in the figure with respect to the phase
transition ($T_{\rm C}\approx 760$\,K). These spectra are not
corrected for temperature factor.}
\end{figure}
\begin{figure}[ht]
\includegraphics[width=8cm]{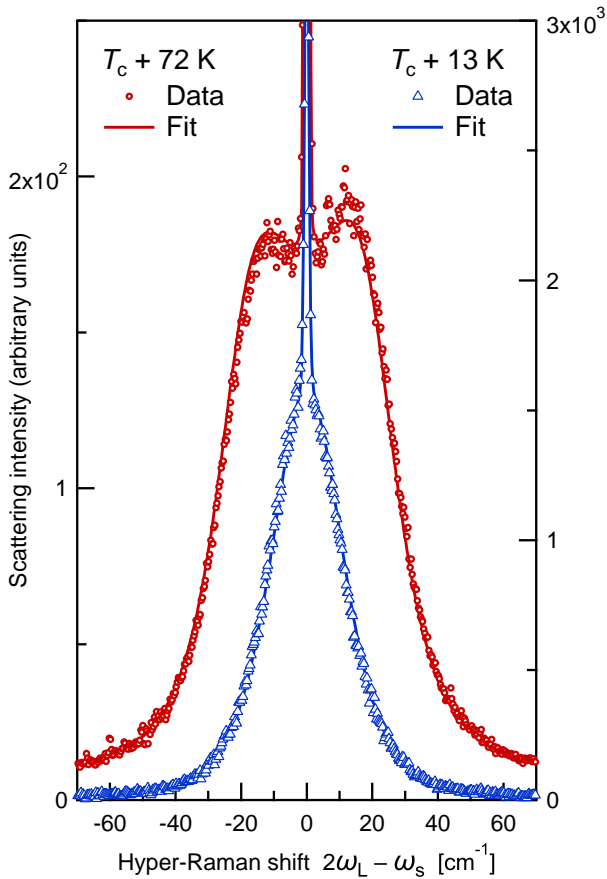}
\caption{(color online) Unpolarized (V+H) HRS spectra of PMN at
13\,K and 72\,K above the phase transition ($T_{\rm C}\approx
760$\,K) adjusted to the model of DHO. The resolution limited
central line is adjusted by a narrow Gaussian peak.
 }
\end{figure}

\begin{figure}[ht]
\includegraphics[width=8cm]{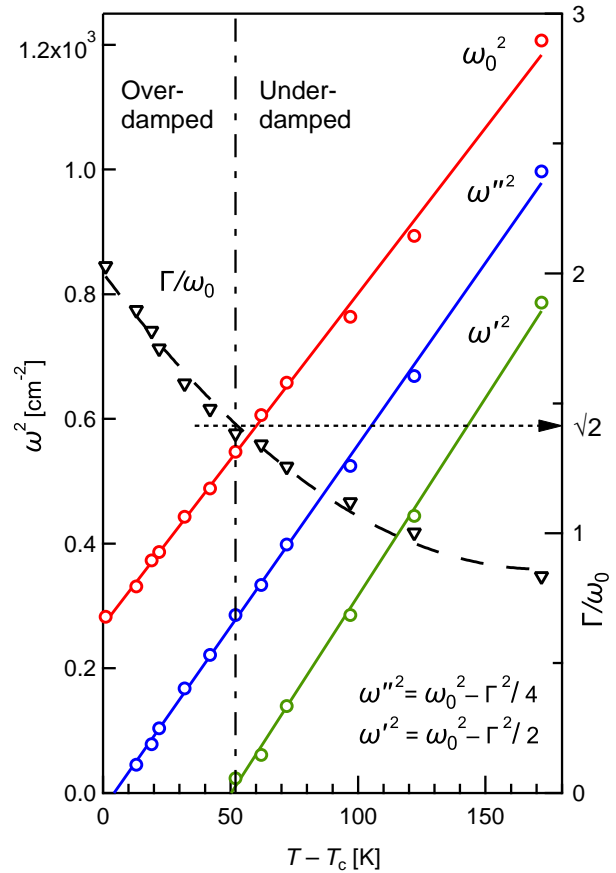}
\caption{(color online) Temperature evolution of soft mode
spectral parameters in paraelectric phase of PbTiO$_3$.
Temperature dependence of the soft mode frequency squared
($\omega_0^2$) was adjusted to a linear (Cochran) law discussed in
the text (full line). The quantities $\omega''^2$ and $\omega'^2$
show analogical behavior. The vertical lines indicated crossover
between overdamped and underdamped soft mode regime. Triangle
symbols stand for the relative soft mode damping - the ratio of
damping and soft mode frequency $\Gamma/\omega_0$. These data
correspond to the right-hand side scale.}
\end{figure}

The narrow central line probably contains both surface sensitive elastic and
 resolution-limited quasielastic scattering, so that we focused here on the analysis of the broad, phonon part of the spectrum. In fact, at all temperatures, this part of the spectrum could be nicely adjusted to the model of a single  DHO, see Fig.\,2.
 Strictly speaking, the measured intensity was adjusted to the sum of a flat background,
  a narrow Gaussian peak standing for the resolution limited central line and the standard one-phonon scattering lineshape model\cite{Vog88}

\begin{equation}
I_{HRS}\propto \left (\frac{\omega_{\rm S}^4}{\omega}\right
)[n(\omega)+1] F(\omega)\sum_\delta |e^{\rm S}_i R_{ijk}^\delta
e^{\rm L}_j e^{\rm L}_k |^2 \, , \label{IHRS}
\end{equation}

where  $\omega_{\rm S} = 2\omega_{\rm L}\pm\omega$ is the
frequency of the scattered field, $n(\omega)$ is the Bose-Einstein
population factor, $\bf{e^L}$ and $\bf{e^S}$ are the polarization
vectors of the incident (L) and detected (S) photons, the index
$\delta$ distinguishes different HRS tensors $R_{ijk}^\delta$ of a
degenerate mode, and $F(\omega)$ is the normalized response
function of the mode described by
\begin{equation}
F(\omega)=(2\omega/\pi)\, {\rm
Im}[1/(\omega_0^2-\omega^2-i\Gamma\omega)]
\end{equation}
where $\omega_0$ and
$\Gamma$ are DHO frequency and damping, respectively.

The resulting least-squared fit values of $\omega_0^2$ as a function of temperature are displayed in Fig.\,3. This temperature dependence is visibly fairly linear, and its adjustment to Cochran law $\omega_0^2= A(T-T_{\rm SM})$ provides  $A= 5$\,cm$^{-2}$K$^{-1}$ and
$T_{\rm SM}=T_{\rm C}-55$\,K. The value of  $\Gamma$ parameter has shown only a very small variation (from about 35\,cm$^{-1}$ at $T_{\rm C}+2$\,K to about 30\,cm$^{-1}$ at $T_{\rm C}+172$\,K).
 Nevertheless,  the value of  $\Gamma$ is comparable with the soft mode frequency itself ($\omega_0$ determined from the present experiments extrapolates to $\sim$17\,cm$^{-1}$ at $T_{\rm C}$).  Therefore, we have also shown in Fig.\,3 the two auxiliary frequency parameters $\omega'=  \sqrt{\omega_0^2-\Gamma^2/2}$ and $ \omega''=
\sqrt{\omega_0^2-\Gamma^2/4}$. The former quantity gives the position of the maximum of the one-phonon DHO scattering spectral
 function of eq.(2) in the high temperature limit ($kT \gg \hbar\omega_{0}$) as long as the DHO is underdamped
 ($\omega_0 > \Gamma/\sqrt{2}$). It is clear from Fig.\,3 that the overdamped/underdamped crossover temperature
$T_{\rm OD}$ at which $\omega'^2$  approaches zero is
 at about 50\,K above $ T_{\rm C}$.

The latter quantity, preferred over $\omega_{0}$ by some
authors,\cite{Soon08} defines the periodic factor in the temporal
response of the oscillator coordinate $Q$ in the absence of
driving forces $Q(t)= \exp^{-t\Gamma /2}\,\cos\left ( \omega''\,t
+ \phi \right )$ for ($\omega_0
> \Gamma/2$). Accidentally,  $\omega'^2$ happens to extrapolate to zero quite close to $ T_{\rm C}$ (see Fig.\,3).


We have noted that the present values of $\omega_0$ reported in
Fig.\,3 fall
 within the error bars of the corresponding values measured  by INS in  a recent paper by Tomeno et al.\cite{Tome12}
  The frequencies deduced from the original work of Shirane  et al\cite{Shir70}
  are a bit higher but still in a reasonable agreement with the present experiment.
 Independent attempt to determine paraelectric soft mode frequency with
  INS technique was reported by Kempa et al.\cite{Kemp06}
Values estimated there implied even higher soft mode frequencies.
We now believe that the main source of differences among these INS
data is the effect of momentum resolution, which was somewhat
coarser in the experiment of Ref.\,\onlinecite{Kemp06}. The soft
mode dispersion of PbTiO$_3$ is indeed very steep,\cite{Tome12} so
that the admixture of the response of the modes with the phonon
wave vectors away from the Brillouin zone center could
considerably upshift the apparent zone center frequency. Another
possibility could be oxygen or lead vacancies in the sample (as
mentioned in Ref.\,\onlinecite{Tome12})
but since the sample used in the present HRS measurements was
prepared under similar conditions as that of the
Ref.\,\onlinecite{Kemp06}, it seems less likely.

The paraelectric soft mode frequency of PTO was also recently
deduced from confocal Raman scattering spectra in
Ref.\,\onlinecite{Soon08}. However, the paraelectric soft mode is
not active in Raman scattering, so that the measured spectral data
did not originate from a standard first-order scattering. The
theoretical analysis of Ref.\,\onlinecite{Soon08}, inspired by
idea of defect-induced Raman scattering is most likely not
appropriate for qualitative determination of the soft mode
parameters at all, as the data obtained there are in a very strong
disagreement with the present measurement - the soft mode
frequencies obtained in Ref.\,\onlinecite{Soon08} are even higher
than that of the INS estimates of Ref.\,\onlinecite{Kemp06}.

The Cochran constant from the present experiments
(5\,cm$^{-2}$K$^{-1}$) has the lowest value among the reported
ones (about 8.4\,cm$^{-2}$K$^{-1}$ is obtained in
Refs.\,\onlinecite{Shir70,Tome12}, and about 35\,cm$^{-2}$K$^{-1}$
is provided in Ref.\,\onlinecite{Soon08}. The latter estimate is
probably biased by the very indirect analysis of the cubic phase
Raman spectra, while the somewhat
higher values reported in Refs.\,\onlinecite{Shir70,Tome12} could
be perhaps related to a broader temperature interval in which the
soft mode was followed (INS data extended up to 1175\,K in
Ref.\,\onlinecite{Tome12}). Such a systematic dependence of the
Cochran constant on the fitting temperature range could indicate a
coupling of the soft mode to a central mode. On the other hand,
the HRS and INS measurements both agree that the Cochran-law
temperature $T_{0}$ is about 50$\pm 15$\,K below the phase
transition $T_{\rm C}$. Present HRS data, taken in the same
temperature region as the dielectric measurements, shows that
$T_{\rm C} - T_{0}$ is close to 55\,K, and so within the
experimental uncertainty, the present Cochran-law temperature
coincides
 with the temperature at which the inverse static susceptibility extrapolates to zero
  (the Curie-Weiss law extrapolating temperature $T_{\rm CW}$ of Ref.\,\onlinecite{Reme70} is 43\,K below $T_{\rm C}$).
   Thus, although the present experiment could not directly probe the  GHz range central peak in paraelectric phase of PTO,
    we can conclude that if it exists, it has only a very small contribution to the static permittivity and also the shift of the phase
     transition is much less than it was sometimes assumed.\cite{Girs97,Kemp06}

 In fact, this simple soft mode picture is also consistent with the observed relative change of the  permittivity at the phase transition.
   Probably the most reliable MHz-range dielectric data of Ref.\,\onlinecite{Li96} suggest that at the phase transition,
    the paraelectric permittivity $\epsilon$ is about 10 times larger than the axial ferroelectric permittivity
    $\epsilon_{33}$ ($\epsilon/\epsilon_{33} \approx 10 $). From the LST relation, assuming that only the soft mode frequency has changed at the transition, the same ratio
could be expressed as $\omega_{A_1}^2/\omega_{\rm SM}^2$, where
$\omega_{A_1}$ is the frequency of the soft mode in the
ferroelectric phase.
 Indeed, the present HRS value of $\omega_{\rm SM}\approx 17$\,cm$^{-1}$ together with the earlier measured\cite{Font91,Jang09}
 values of $\omega_{A_1}\approx 60$\,cm$^{-1}$ leads us also to the ratio of  $\omega_{A_1}^2/\omega_{\rm SM}^2 \approx 10$.
 In summary, we can conclude that the PbTiO$_3$ can be considered as a model soft-mode driven ferroelectric system.

\begin{acknowledgments}
This work has been supported by the Czech Ministry of Education
(Project MSMT ME08109).
\end{acknowledgments}

\bibliographystyle{plain}

\end{document}